\documentclass[aps,twocolumn,prl,
groupedaddress,nofootinbib,nobalancelastpage,nobibnotes]{revtex4}
\pdfoutput=1
\usepackage{amsmath,amsfonts,amssymb,mathrsfs,graphicx}
\usepackage[squaren]{SIunits}

\newcommand{\ie}{{\it i.e.}}

\newcommand{\eg}{{\it e.g.}}

\newcommand{\cf}{{\it cf.}}

\newcommand{\eq}{Eq.}

\newcommand{\fig}{Fig.}

\newcommand{\Ref}{Ref.}
\newcommand{\Refs}{Refs.}

\newcommand{\equ}[1]{\eq~(\ref{equ:#1})}
\newcommand{\figu}[1]{\fig~\ref{fig:#1}}

\newcommand{\bi}{\begin{itemize}}
\newcommand{\ei}{\end{itemize}}

\begin{document}

\title{Neutrino Emission from Gamma-Ray Burst Fireballs, Revised}

\author{Svenja H{\"u}mmer}
\affiliation{Institut f{\"u}r theoretische Physik und
  Astrophysik, Universit{\"a}t W{\"u}rzburg, Am Hubland, D-97074 W{\"u}rzburg, Germany}

\author{Philipp Baerwald}
\affiliation{Institut f{\"u}r theoretische Physik und
  Astrophysik, Universit{\"a}t W{\"u}rzburg, Am Hubland, D-97074 W{\"u}rzburg, Germany}

\author{Walter Winter}
\affiliation{Institut f{\"u}r theoretische Physik und
  Astrophysik, Universit{\"a}t W{\"u}rzburg, Am Hubland, D-97074 W{\"u}rzburg, Germany}

\date{\today}

\begin{abstract}
We review the neutrino flux from gamma-ray bursts, which is estimated from gamma-ray observations and used for the interpretation of recent IceCube data, from a particle physics perspective. We numerically calculate the neutrino flux for the same astrophysical assumptions as the analytical fireball neutrino model, including the dominant pion and kaon production modes, flavor mixing, and magnetic field effects on the secondary muons, pions, and kaons. We demonstrate that taking into account the full energy dependencies of all spectra, the normalization of the expected neutrino flux reduces by about one order of magnitude and the spectrum shifts to higher energies, where we can pin down the exact origin of the discrepancies by the re-computation of the analytical models. We also reproduce the IceCube-40 analysis for exactly the same bursts and same assumptions and illustrate the impact of uncertainties. We conclude that the baryonic loading of the fireballs, which is an important control parameter for the emission of cosmic rays, can be constrained significantly with the full-scale experiment after about ten years.
\end{abstract}

\maketitle

If gamma-ray bursts (GRBs) are sources of ultra-high energy cosmic rays (UHECRs), they should also lead to neutrino production~\cite{Waxman:1997ti,Waxman:1998yy}. The IceCube~\cite{Ahrens:2003ix} (IC) neutrino telescope has, for the first time, significantly constrained the neutrino flux to below the expectations from gamma-ray and cosmic ray observations~\cite{Abbasi:2011qc,Abbasi:2012zw}; see also \Ref~\cite{Ahlers:2011jj} for a fit to cosmic ray data. In particular, the method to compute the expected neutrino flux from the gamma-ray fluence in the internal shock model has been derived in \Refs~\cite{Guetta:2003wi,Abbasi:2009ig}; we refer by ``IC Fireball (neutrino) Calculation'' (IC-FC) exactly to the IceCube version in \Ref~\cite{Abbasi:2009ig}, which is based on \Ref~\cite{Guetta:2003wi}. On the other hand, it has been clear from numerical calculations that there are limitations to the analytical method from the particle physics perspective, see, for instance, \Refs~\cite{Murase:2005hy,Kashti:2005qa,Asano:2006zzb,Lipari:2007su,Hummer:2010vx,Baerwald:2010fk,Baerwald:2011ee} for the impact of additional pion/kaon production modes, flavor mixing, and magnetic field effects on the secondary muons, pions, and kaons. 
For example, normalizing the proton and photon densities in the source to the Waxman-Bahcall (WB) GRB flux~\cite{Waxman:1998yy}, it has been demonstrated in \Ref~\cite{Baerwald:2010fk} that the combination of these effects modifies the shape significantly, and increases the normalization by a factor of three to four. So obviously there has been increasing tension between theory and observation, which has challenged the paradigm that GRBs are the sources of the UHECR. We study
the connection between gamma-ray observations and neutrinos  by re-interpreting the IC40 data with a numerical model based on exactly the same assumptions, same parameters, and same bursts, \ie, without changing the astrophysical ingredients. However, we include the additional multi-pion, kaon, and neutron production modes, the synchrotron losses of the secondaries, adiabatic cooling, and the full energy dependence of the spectra. Compared to \Ref~\cite{Baerwald:2010fk}, we do not normalize the neutrino flux to the WB flux, but to the actually observed photon fluence. That is, the photon density in the source is obtained from $E_\gamma^{\mathrm{iso}}$ following the gamma-ray observation, the magnetic field is obtained from energy equipartition between electrons and magnetic field, and the proton density is assumed to follow an $E^{-2}$ injection spectrum with the normalization determined by the baryonic loading. Since we find significant discrepancies in normalization and shape compared to the analytical models, we pin down the differences by re-computations of the (original) analytical models in \Refs~\cite{Waxman:1997ti,Guetta:2003wi,Abbasi:2009ig}. 

\begin{figure*}[t!]
\begin{center}
\includegraphics[width=0.35\textwidth]{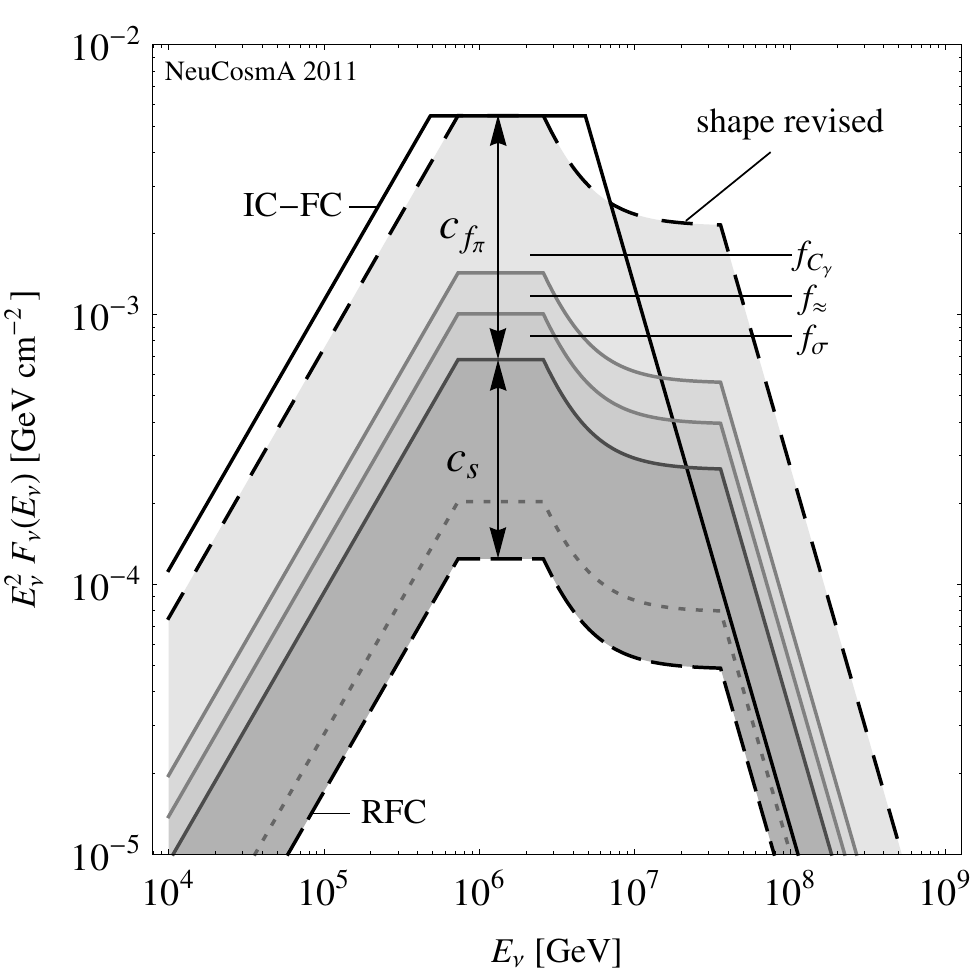} \hspace*{0.5cm}
\includegraphics[width=0.35\textwidth]{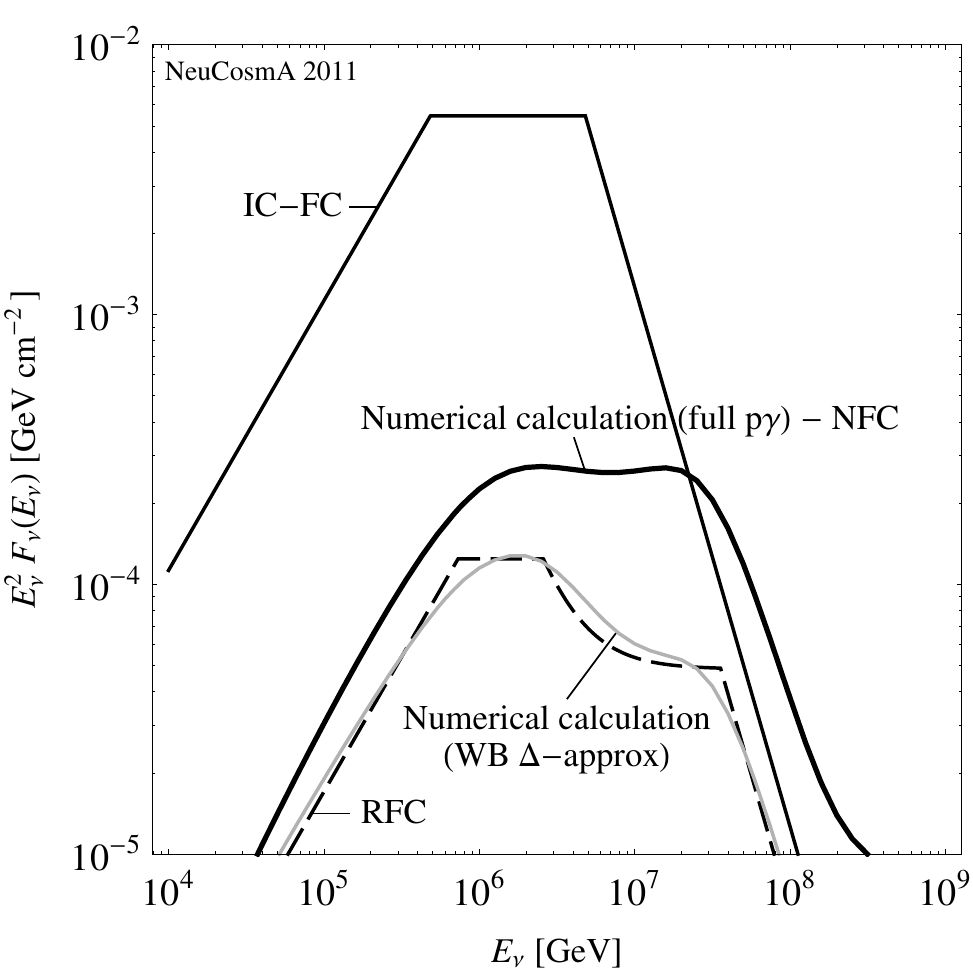}
\vspace*{-0.5cm}
\end{center}
\caption{\label{fig:rfb} Left panel: Shape and normalization modification for the muon neutrinos from GRB080603A between IC-FC and the Revised (analytical) Fireball Calculation (RFC), where the individual contributions are described in the main text (burst parameters: photon spectrum with $\alpha=1$, $\beta=2$, $\epsilon^b_\gamma = 0.2 \, \mathrm{MeV}$, fluence $\mathcal{F}_\gamma = 5.5\cdot10^{-5} \, \mathrm{erg\,cm^{-2}}$, $\Gamma = 10^{2.5}$, $t_v =0.01 \, \mathrm{s}$, $T_{90} =180 \, \mathrm{s}$, $z = 1.69$,  $L_\gamma^{\mathrm{iso}}=6.0\cdot10^{51} \, \mathrm{erg \, s^{-1}}$~\cite{PC}; $B \simeq 90 \, \mathrm{kG}$, as a consequence of the assumed energy equipartition fractions and parameters; see Eq.~(17) of \Ref~\cite{Baerwald:2011ee}). Right panel: Numerical  reproduction of RFC using the WB $\Delta$-resonance approximation~\cite{Waxman:1997ti}, and the finally obtained calculation NFC including the additional pion and kaon production modes.   }
\end{figure*}

First of all, consider the analytical method IC-FC in \Ref~\cite{Abbasi:2009ig} (see App.~A therein), used for the IceCube analyses. At the source, protons, injected with an $E^{-2}$ spectrum, are assumed to collide with target photons with a broken power law spectrum which comes from the gamma-ray observation on a burst-by-burst basis.  The neutrino spectrum is assumed to have two breaks, one from the photon spectrum, and one from the cooling of the secondaries. The normalization of the neutrino fluence is computed from the photon fluence with
\begin{align}
  \int_0^\infty dE_\nu\, E_\nu F_\nu(E_\nu) 
& = \frac{1}{8} \underbrace{\left(1-\left(1- \langle x_{p \rightarrow \pi} \rangle \right)^{\Delta R/\lambda_{p\gamma}}\right)}_{f_\pi} \nonumber \\ 
& \times \frac{1}{f_e}\int_{1\,\text{keV}}^{10\,\text{MeV}}d\varepsilon_\gamma\, \varepsilon_\gamma \, F_\gamma(\varepsilon_\gamma) \,  \label{equ:normIC} 
\end{align}
where  $\langle x_{p \rightarrow \pi} \rangle \simeq 0.2$ is the (average) fraction of proton energy going into a pion per interaction, $f_e$ is the fraction of the total energy in electrons compared to the total energy in protons ($1/f_e$: baryonic loading), $f_\pi$ is the pion production efficiency, $\lambda_{p \gamma}=1/(n_\gamma \sigma_\Delta)$ is the proton mean free path, and $\Delta R$ is the shell width. On the other hand, our Numerical Fireball Calculation (NFC) is described in detail in \Ref~\cite{Baerwald:2011ee} (model ``FB-D'' therein). In short, the model relies on the proton and photon densities within the source. Once these spectra are fixed, the rest is just particle physics, where the effect of synchrotron cooling, adiabatic cooling, and decay of the secondaries is explicitely computed (compared to the analytical approach). The normalization of the photon density is obtained by calculating the equivalent energy from the measured photon fluence 
\begin{equation}
 E_\gamma^{\mathrm{iso}}= \frac{4 \pi d_L^2}{1+z} \, \int_{1\,\text{keV}}^{10\,\text{MeV}}d\varepsilon_\gamma\, \varepsilon_\gamma \, F_\gamma(\varepsilon_\gamma) \, .
\label{equ:eiso}
\end{equation} 
Then the proton and magnetic (energy) density normalizations are obtained by the usual energy partition assumptions with the same parameters as in \Refs~\cite{Abbasi:2009ig,Abbasi:2011qc}, and the same assumptions for the geometry of the fireball. Note that the IceCube analysis is based on a number of bursts for which the neutrino flux is ``stacked'', since the expected neutrino signal from one burst is too small. Therefore, this computation has to be performed for each burst individually.

Let us now compare the results of IC-FC and NFC by using a simplified analytical version of the numerical code, based on the photo-meson production in \Ref~\cite{Waxman:1997ti} (``WB $\Delta$-approx''), and by re-computing the analytical models. The main difference has been identified to be spectral effects: while the analytical computations often rely on estimates using a particular energy (\eg, the photon break energy), the numerical code takes into account the full energy dependencies automatically. In \figu{rfb}, we illustrate this in the left panel for one specific example from the IC40 analysis producing a result similar to \Refs~\cite{Waxman:1997ti,Waxman:1998yy}. The curve IC-FC shows the analytical expectation for the chosen parameter set. As a first step, the shape is revised (curve ``shape revised'') by including a shift of the first break (correction of threshold of photohadronic interactions in Eq.~(A3) of \Ref~\cite{Guetta:2003wi}, see \Ref~\cite{Baerwald:2010fk}, or missing factor in Eq.~(3) of \Ref~\cite{Waxman:1997ti}), the fact that there are two different cooling breaks for muons and pions (including flavor mixing), and a factor of $1+z$ from the effect of the cosmic expansion on the variability timescale. As the next step, the correction $c_{f_\pi}$ to the pion production efficiency contains: $f_{C_\gamma}$ (energy of all photons approximated by break energy, whereas photons distributed according to the photon spectrum; coming from Eq. (A13) in~\cite{Guetta:2003wi}), $f_{\approx}=0.69$ (rounding error in Eq. (A15) in~\cite{Guetta:2003wi}), and $f_\sigma \simeq 2/3$ (from neglecting the width of the $\Delta$-resonance in $\lambda_{p \gamma}$ instead of using the interaction rate; after Eq. (A12) in~\cite{Guetta:2003wi}, but included in Eq.~(3) of \Ref~\cite{Waxman:1997ti}). The factor $c_S$ corrects for energy losses of the secondaries and the energy-dependence of the mean free path of the protons, see Eq.~(11) in \Ref~\cite{Li:2011ah} and discussion therein. 
Note that this factor is somewhat model-dependent because the energy in protons, computed from the energy partition and baryonic loading, depends logarithmically on the minimal and maximal proton energies (for an $E^{-2}$ injection spectrum),  whereas in \equ{normIC} only the part relevant for neutrino production is taken into account. To illustrate that, the  dotted curve in \figu{rfb}, left, shows the result in the extreme case that the minimal proton energy coincides with the photo-meson production threshold. Note that $f_{C_\gamma}$ and $c_S$ strongly depend on the photon spectral indices, and vary from burst to burst.  Surprisingly, all the corrections go into the same direction, which means that the approximations in the analytical model were probably a bit on the optimistic side.  The result can be regarded as Revised Fireball (neutrino) Calculation (RFC).
In the right panel of \figu{rfb}, it is shown that RFC matches the numerical result for the same assumptions for the photohadronic interactions (``WB $\Delta$-approx'', as in \Ref~\cite{Waxman:1997ti}) very well. Taking into account the additional multi-pion and kaon production modes, similar to \Ref~\cite{Baerwald:2010fk}, the flux increases again, and the final (numerical) result NFC is obtained. In this case, the normalization deviates about one order of magnitude from the analytical prediction IC-FC, and the shape is significantly different, shifted to higher energies.  Note that we have chosen one analytical method IC-FC for the comparison, whereas the detailed comparison to another method, such as \Ref~\cite{Waxman:1997ti}, will depend on the specific approximations of the analytical method (whereas NFC does not depend on these).

\begin{figure}[tb]
\begin{center}
\includegraphics[width=0.8\columnwidth]{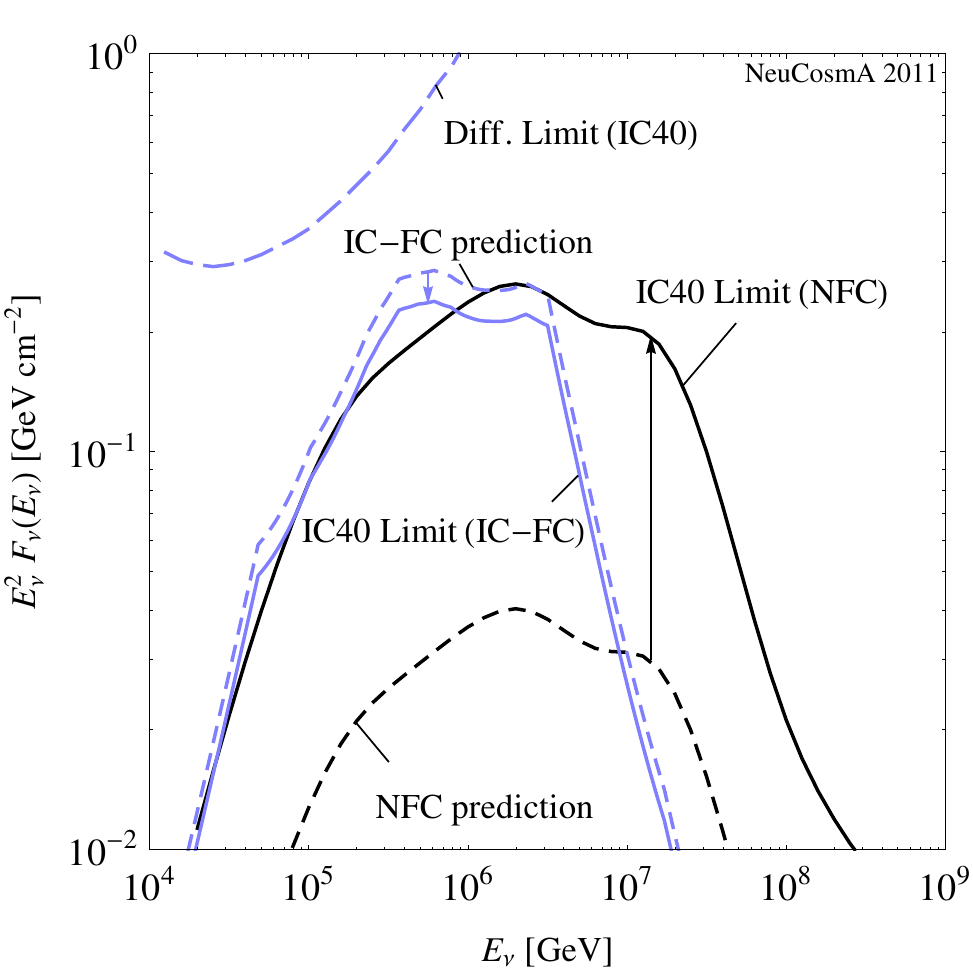}
\vspace*{-0.5cm}
\end{center}
\caption{\label{fig:limit} Reproduction of the IC-FC prediction for the neutrino (differential) fluence $E_\nu^2 F_\nu$, compared to the  corresponding IC40 limit (light/blue curves; 90\% CL). In addition, our numerical prediction NFC is compared to the corresponding IceCube limit for exactly the same bursts and assumptions (black curves). Compare to Fig.~2 in \Ref~\cite{Abbasi:2011qc}.}
\end{figure}

As the next step, we reproduce the IC40 analysis from \Ref~\cite{Abbasi:2011qc}, based on 117 bursts, using the same neutrino effective area and same assumptions, bursts, and parameters~\cite{PC}. The result is shown in \figu{limit} (light/blue curves), where the dashed curve shows the IC-FC prediction for the neutrino flux and the solid curve the corresponding IC40 limit. In this case, the bound is below the prediction, and the original model is under tension. Our result is shown as black curves: the prediction is about one order of magnitude below the limit corresponding to this flux shape. This qualitatively different result means that IceCube has not yet reached the level where it tests the parameters chosen for the fireball model.

\begin{figure}[b!]
\begin{center}
\includegraphics[width=0.8\columnwidth]{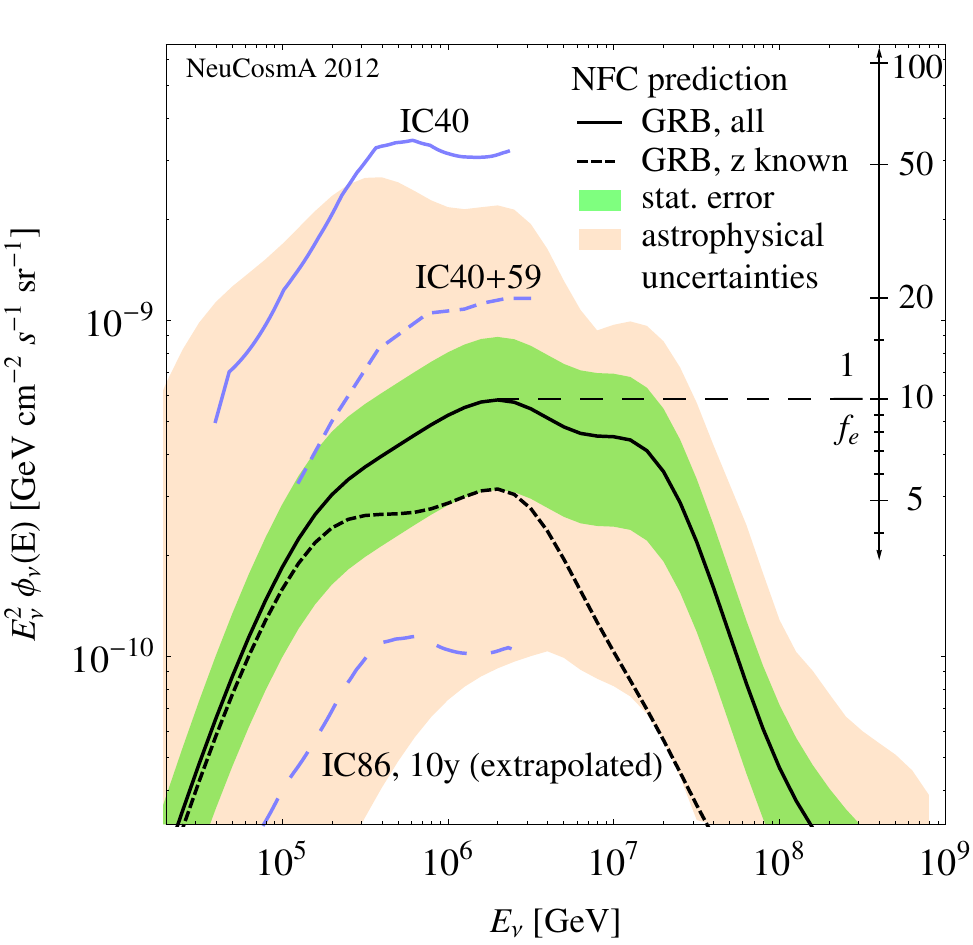}
\vspace*{-0.5cm}
\end{center}
\caption{\label{fig:syst} Prediction of the quasi-diffuse flux (NFC), including the estimates for several model- or method-specific systematical uncertainties (see main text). In addition, the IC40 limit is shown, and two expectations are shown for comparison (IC59+40 from \Ref~\cite{Abbasi:2012zw} and IC86 extrapolated  for $A_{\mathrm{eff}}^{\mathrm{IC86}} \simeq 3 \times A_{\mathrm{eff}}^{\mathrm{IC40}}$ from IC40; see, \eg, \Ref~\cite{Karle:2010xx}).}
\end{figure}

In order to obtain conclusions on the cosmic-ray connection, or to compare the results from different experiments, the extrapolation of the fluence to a quasi-diffuse flux is needed. It depends on the number of bursts expected per year, where 667 has been used~\cite{Abbasi:2011qc}. We show in \figu{syst} our quasi-diffuse flux prediction (``GRB, all'') together with the IC40 limit, the combined IC59+40 limit (which has a different flux shape), and an extrapolated IC86 limit. In addition, we show different regions and curves to illustrate the size of several model- or method-specific additional ``systematical errors'': the statistical error coming from the extrapolation from a few bursts to the quasi-diffuse flux (for 117 bursts, estimated and obtained from \Ref~\cite{Baerwald:2011ee}) and the ``astrophysical uncertainty'' for this particular model (envelope of the
following independent variations around the assumptions for the IceCube analysis: variability timescale $t_v$ by one order of magnitude [$0.001 \mathrm{s} \hdots 0.1 \mathrm{s}$  for long bursts], $\Gamma$ between 200 and 500, proton injection index between 1.8 and 2.2, and $\epsilon_e/\epsilon_B$, energy in electrons versus magnetic field, between 0.1 and 10). 
As one can read off from this figure, neither IC40 nor IC59+40 can reach the predicted fluxes, even in the most optimistic cases; compared to IC59+40, a factor of two higher statistics is needed to reach the nominal prediction. However, the full scale IceCube experiment, operated over about 10~years (extrapolation), will finally find the GRB neutrinos or significantly constrain the model unless, for instance, the number ratio between $\Gamma \gtrsim 500$ and $\Gamma \sim 300$ bursts (or corresponding collision radii) is larger than seven for fixed $t_v$, as it can be easily shown.  Note that our given astrophysical uncertainty is less model-dependent than the one in \Ref~\cite{Guetta:2001cd}, since it does not rely on the origin of the target photons, but it includes the effects of synchrotron losses.

We have deliberately omitted one variable from this discussion: the baryonic loading $1/f_e$, which directly re-scales the neutrino flux prediction, as illustrated by the arrow in \figu{syst} and as it can be read off from \equ{normIC}. The choice of this parameter is often consistent with a coherent picture among cosmic ray, gamma-ray, and neutrino fluxes if the GRBs are the sources of the UHECR, \ie, treating it at the same level as the other parameters would change the logic of the astrophysical picture. 
For example, if the protons are magnetically confined within the source, the cosmic ray flux will be proportional to baryonic loading and the photohadronic interaction rate, and therefore to the neutrino flux. 
As a consequence, a limit on $1/f_e$ can be translated into a limit on the cosmic ray production in specific models, and may even challenge the paradigm that GRBs are the sources of the UHECR; see, \eg, \Ref~\cite{Ahlers:2011jj}. It can be read off from \figu{syst} that IC59+40 already constrains the baryonic loading to below about~20. Future IceCube data may reach $1/f_e \simeq 2$, which, in turn, means that about one third of the energy of the GRB goes into photons. If this fraction is unacceptably large for theoretical GRB models, they will be finally challenged. 

There are, however, some caveats in this analysis: For most of the bursts used for the IC40 analysis the redshift is actually not measured. In these cases, for long bursts in \Ref~\cite{Abbasi:2011qc}, $z=2.15$ has been used for the computation of the breaks, and $L_\gamma^{\mathrm{iso}}=10^{52} \, \mathrm{erg \, s^{-1}}$ for the computation of $f_\pi$ in \equ{normIC}. However, in a self-consistent numerical approach, it is clear that these two quantities are correlated (\cf, \equ{eiso} for $F_\gamma$ measured, with $L_\gamma^{\mathrm{iso}} \simeq (1+z) E_\gamma^{\mathrm{iso}}/T_{90}$). We have in \figu{limit} followed the IceCube logic in these cases: we have computed the proton density from $F_\gamma$ directly, and the photon density from the chosen $L_\gamma^{\mathrm{iso}}$ (as it enters $f_\pi$) -- using the chosen value of $z$ in both cases. Note, however, that fixing $z$ and computing the luminosity from $F_\gamma$ with \equ{eiso}  in a self-consistent approach leads to a prediction which is strongly dependent on the chosen ``standard value'' of $z$. In this case, choosing $z \simeq 2$ clearly overestimates the neutrino flux, where we find that one burst with a large photon fluence dominates. On the other hand, the peak contribution to the diffuse flux may rather come from $z \simeq 1$~\cite{Baerwald:2011ee}.
Because of this uncertainty, a conservative lower limit on the neutrino flux can be only obtained from the bursts with $z$ measured, \cf, lower dashed curve in \figu{syst}. A similar logic could be applied to $t_v$ and $\Gamma$.

In summary, we have revised the GRB fireball neutrino flux calculation from a particle physics perspective. Compared to earlier analytical computations, our numerical simulation takes into account the full spectral (energy) dependencies of the proton and photon spectra, as well as the cooling of the secondaries, flavor mixing, and additional multi-pion, kaon, and neutron production channels for the neutrinos. We have found a significant deviation in the normalization of the neutrino flux prediction of about one order of magnitude, with a very different spectral shape peaking at slightly higher energies. We have shown from the re-computation of the analytical models where the discrepancies come from, and we have demonstrated that they all add up in the same direction. Note that the exact size of the corrections depends on the individual burst parameters, which we have taken into account. 
We have also demonstrated, by the reproduction of the IC40 analysis, that our prediction is significantly below the current IceCube limit, which means that the conventional GRB fireball phenomenology is not yet challenged. Finally, we have quantified additional astrophysical and method-dependent systematical uncertainties in the computation for this particular model in order to determine a lower bound for the predicted neutrino flux. We have demonstrated that the baryonic loading of the fireballs can be constrained with the full-scale experiment after about ten years to a level which will exert significant pressure on the parameter space for GRBs as sources of the UHECR.
Note, however, that we have only considered the simplest possible one-zone internal shock model, whereas the spectral shape may change in multi-zone models (see, \eg, \Ref~\cite{Li:2008ub}) or taking into account additional effects, such as the acceleration of the secondaries~\cite{Koers:2007je}. In addition, photospheric emission and magnetic reconnection models may lead to larger emission radii, see, \eg, \Ref~\cite{Murase:2008sp}, and UHECR acceleration may also be possible at external shocks~\cite{Waxman:1999ai,Dermer:2000yd} -- with potentially smaller neutrino fluxes.


We would like to thank M. Ahlers, D. Guetta, A. Kappes, M. Richman, and E. Waxman for useful discussions and comments, and K. Meagher and N. Whitehorn for help with the reproduction of the IC40 analysis. 
 PB and SH acknowledge support  from GRK1147 of DFG, SH from the Studien\-stif\-tung des deutschen Volkes, and WW from DFG contracts WI 2639/3-1 and WI 2639/4-1.


\end{document}